\documentclass[%
 reprint,
amsmath,amssymb,
aps,
]{revtex4-2}

\usepackage[english]{babel}
\usepackage[utf8]{inputenc}
\usepackage{amsmath}
\usepackage{physics}
\usepackage{graphicx}
\usepackage[colorinlistoftodos]{todonotes}
\usepackage{lipsum}
\usepackage{multirow}
\usepackage{calc}
\usepackage[bottom]{footmisc}
\usepackage[]{hyperref}
\usepackage{booktabs}
\usepackage[normalem]{ulem}
\usepackage{float}
\usepackage{lineno}
\usepackage{xcolor}
\usepackage[normalem]{ulem}

\usepackage{xcolor}
\usepackage{braket}
\usepackage{subfig}

\begin{document}

\title{Low-noise quantum frequency conversion in a monolithic cavity with bulk periodically poled potassium titanyl phosphate}

\author{Felix Mann$^{1,*}$, Helen M. Chrzanowski$^{1}$, Felipe Gewers$^{2}$, Marlon Placke$^{1}$, Sven Ramelow$^{1,3}$}

\affiliation{$^{1}$\mbox{Institut f\"ur Physik, Humboldt-Universit\"at zu Berlin, Newtonstr. 15, 12489 Berlin, Germany} \\
$^{2}$\mbox{Instituto de Física, Universidade de São Paulo, R. do Matão 1371, São Paulo 05315-970, Brazil}\\
$^{3}$\mbox{IRIS Adlershof, Humboldt-Universität zu Berlin, Berlin, Germany}\\
$^{\star}$\mbox{Corresponding author: felixmann@physik.hu-berlin.de}}

\begin{abstract}

Interfacing the different building blocks of a future large scale quantum network will demand efficient and noiseless frequency conversion of quantum light. Nitrogen-vacancy (NV) centers in diamond are a leading candidate to form the nodes of such a network. However, the performance of a suitable converter remains a bottleneck, with existing demonstrations severely limited by parasitic noise arising at the target telecom wavelength. Here, we demonstrate a new platform for efficient low-noise quantum frequency conversion based on a monolithic bulk ppKTP cavity and show its suitability for the conversion of 637 nm single photons from NV centers in diamond to telecommunication wavelengths. By resonantly enhancing the power of an off-the-shelf pump laser, we achieve an internal conversion efficiency of $(72.3\pm 0.4) \%$ while generating only $(110\pm 4) \mbox{ kHz/nm}$ noise at the target wavelength without the need for any active stabilization. This constitutes a 5-fold improvement in noise over existing state-of-the-art single-step converters at this wavelengths. We verify the almost ideal preservation of non-classical correlations by converting photons from a spontaneous parametric down-conversion source and moreover show the preservation of time-energy entanglement via Franson interferometry.

\end{abstract}

% Ions \cite{rempe} dots \cite{lodahl}.

\maketitle

%%----------INTRODUCTION------------------------------------------------------------------------

\section{Introduction}

Quantum frequency conversion (QFC), the coherent interchange of quantum states between light beams of different frequencies \cite{kumar}, will be indispensable for interfacing the different building blocks of future heterogeneous quantum networks \cite{kimble,awschalom}. Several platforms are currently extensively investigated for the implementation of such networks: trapped ions \cite{ions}, atoms \cite{rempe}, semiconductor quantum dots \cite{lodahl} and color centers in diamond \cite{hanson_color}. Here specifically nitrogen-vacancy (NV) centers in diamond are a promising platform \cite{hansonNV,munroNV}. This is due to the long coherence time of the electronic spin associated with this defect center \cite{nv_coherence}, the possibility to couple this electronic spin to nuclear spin \cite{register,nv_memory,nv_memory2} and an optical interface at 637 nm \cite{wrachtrup}. Thus NV centers in diamond have the potential of being used to process, store and transmit quantum information \cite{hanson3}. 
\\Without the conversion of the NV center's 637 nm single photons to the telecommunication band, they suffer from strong transmission losses in optical fiber, severely limiting the suitability of NV centers for large scale quantum networks \cite{dreau}. As a potential solution, efficient QFC of single photons has been demonstrated using sum- or difference-frequency generation (SFG, DFG) driven by a strong pump laser. QFC by SFG or DFG is usually realized in periodically-poled $\chi^{(2)}$-nonlinear crystals, with the conversion efficiency $\eta_c$  as a function of pump power $P_P$ given by \cite{albota}  
\begin{gather}
\begin{split}
\eta_c = \sin^2\left( \frac{\pi}{2}\sqrt{\frac{P_p}{P_{max}}}\right),\\ 
P_{max} = \frac{c \epsilon_0 n_t n_r \lambda_t \lambda_r \lambda_p}{128d^2_{eff}L h_m(\xi_p, \xi_r)},
\label{eq:DFG}    
\end{split}
\end{gather}
where $\lambda _x$ is the vacuum wavelength with the subscripts denoting $t$ for the target wavelength (or telecom), $r$ for the input "red" wavelength and $p$ for the pump wavelength, and  $n_x$ denoting their corresponding refractive indices. Further, $d_{eff}$ is the nonlinear coefficient, $L$ is the length of the nonlinear crystal, and $h_m$ the reduction factor for focused Gaussian beams, which is a function of the focusing parameters $\xi_x$ of the participating beams \cite{bk,guha}.
\\The central challenge for all implementations of frequency converters - for which ppLN waveguides have historically been the favoured architecture - is noise arising at the target conversion wavelength \cite{albota,dreau,ikuta,strassmann,maring}. The dominant noise processes are typically Raman scattering and fluorescence of the pump and, in situations where the pump wavelength is shorter than the target wavelength, parasitic spontaneous parametric down-conversion (SPDC) driven by errors in the period poling \cite{pedestal,NSD,fejer92}. The efficient upconversion of this parasitic SPDC also limits the performance of converters where the target wavelength is shorter than the pump wavelength, but where the pump is located in between the two wavelengths which are converted \cite{albota}. This applies to most conversion scenarios which interconnect visible and telecommunication wavelengths. Further examples of potential building blocks of quantum networks which would benefit from a visible to telecom interconnect, that are affected by parasitic SPDC, are other color centers in diamond like GeV (602 nm), SnV (620 nm), PbV (520/555 nm) \cite{hanson_color}, quantum memory at 606 nm \cite{memory606} and 580 nm \cite{memory580}, trapped ions (369.5 nm) \cite{ions2} but also hexagonal boron nitride (500 nm-850 nm, mainly around 600 nm) \cite{hBN_1,hBN_2}. One approach to decrease the noise caused by SPDC is a two-step conversion proposed by \cite{albota} and demonstrated by \cite{fejer2step}. A drawback of this approach is the increased complexity of the optical system, hindering their application readiness for larger-scale networks.
\\In our previous work, it was hypothesised that bulk ppKTP could prove a promising platform for quantum frequency conversion, owing to the comparatively high quality of the periodic poling in bulk KTP \cite{carlota,kthdomain,mutter}  which may help suppress parasitic SPDC processes \cite{mann}. However, waveguide converters based on ppKTP also suffer from significant parasitic SPDC noise \cite{ppKTPWGnoise}, which can potentially be attributed to inhomogeneities in the waveguide geometry and a Cerenkov-idler configuration \cite{rastogi,NSD} yielding non-negligible phase-matching for parasitic SPDC processes.\\
Here, we demonstrate the potential of bulk ppKTP for efficient low-noise quantum frequency conversion from 637 nm to telecommunication wavelengths. But equally, these results can be applied to all conversion scenarios where the pump lies between the wavelengths which are converted. To obtain the necessary pump power for efficient conversion in a bulk crystal, the pump power was enhanced via a cavity. In particular, we did employ a monolithic cavity design to circumvent the requirement for active stabilisation \cite{mitchell,mitchell2}. This passively stable, resonant enhancement of the pump enabled the use of a low-cost off-the-shelf 1064 nm continuous-wave (CW) pump laser with a power of 3 W, without introducing additional experimental complexity.\\
\section{Experimental setup} 
\begin{figure}[!htbp]
\centering
\includegraphics[width=8.8cm]{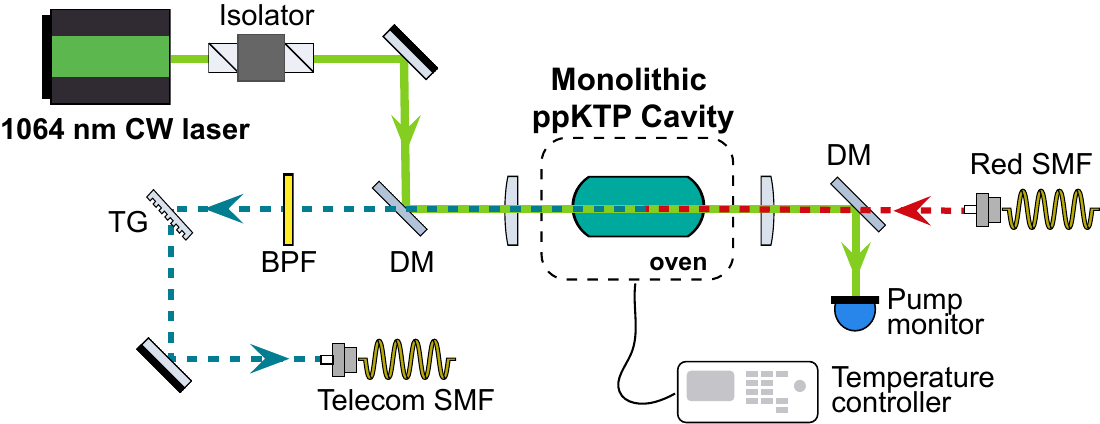}
\caption{Frequency conversion setup with the monolithic bulk ppKTP cavity: A CW 1064 nm pump laser with 3 W power was coupled into the monothic ppKTP cavity. For conversion from 637 nm to the telecom band, the red light was launched from a single mode fiber (SMF) backwards into the ppKTP crystal, where it underwent difference frequency generation with the back-travelling pump wave, generating light at the target telecom wavelength. The resulting telecom light was spectrally filtered via a long-pass dichroic mirror (DM), a bandpass filter (BPF, FWHM = 50 nm) and a holographic-transmission grating (TG). The transmitted light was then coupled into a SMF and subsequently detected.}
\label{fig:setup}
\end{figure}

The monolithic bulk ppKTP cavity was made from a $L=20 \mbox{ mm}$ long KTP crystal (Raicol Crystals Ltd.) quasi-phase matched for the type-0 DFG (SFG) process 637 nm $\rightarrow$ 1064 nm + 1587 nm (or vice versa) with a domain length of $l = 7.85\mbox{ }\mu\mbox{m}$. The conversion bandwidth is 110 GHz (FWHM) corresponding to 0.9 nm (0.15 nm) at 1587 nm (637 nm).
\\In order to obtain a stable pump enhancement cavity, both end facets were polished spherically convex with a radius of curvature of $\mathcal{R}=14\mbox{ mm}$ and coated to be anti-reflective for the red/telecom beams and highly reflective ($R_{p}=98\%$) for the 1064 nm pump. $\mathcal{R}$ defines the cavity/pump mode to have a focusing parameter of $\zeta_p \approx 1.55$ which, after matching the focusing parameters for the red/telecom modes, leads to a Gaussian beam reduction factor of $h_m\approx 0.9$ \cite{bk,guha}. We measured the finesse of the cavity to be $\mathcal{F}= 146$ which is in agreement the predicted finesse within the fabrication tolerances. The power enhancement of $\mathcal{E}=50$ resulted in a maximum circulating power of about 75 W when pumped with 3W at 1064 nm. This 75 W corresponds to a coupling efficiency of 65\% into the fundamental mode, inferred by power measurements behind the cavity. With an appropriate housing the temperature was stabilized to below $1 \mbox{ mK}$ at room temperature. The full experimental details of the conversion setup are provided in Figure~\ref{fig:setup}.\\
The noise floor of the converter was measured with a commercial superconducting nanowire single-photon detector (SNSPD) system (Quantum Opus LLC). Each channel had a quantum efficiency of $\eta_{det}\approx$ 90\% and a dark count rate of approximately 100 Hz.
\section{Results}
\begin{figure}[!htbp]
\centering
\includegraphics[width=7.3cm]{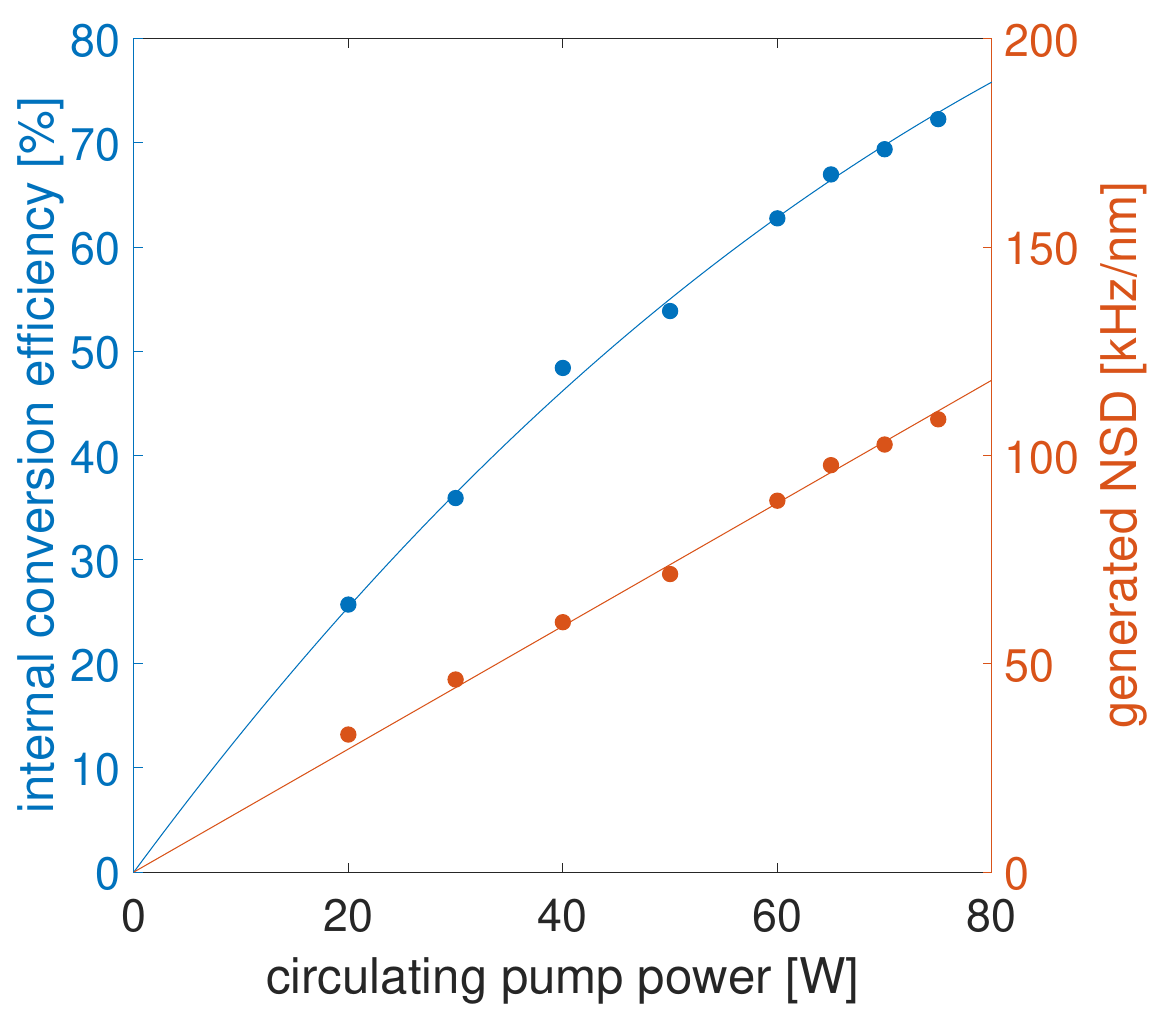}
\caption{Measured internal conversion efficiency (blue dots) and the generated noise spectral density (NSD) (orange dots) versus circulating pump power. The internal conversion efficiency was determined by measuring the depletion of a weak ($\sim100$ $\mu $W) 637 nm probe laser. At the maximum circulating pump power of $(74.5\pm 0.3)\mbox{ W}$ the conversion efficiency reached $(72.3\pm 0.4)\%$. The conversion efficiency was fitted with  $\sin^2(\pi/2\sqrt{P_p/P_{max}})$ with the fit predicting 100\% internal efficiency at $P_{max}=177 \mbox{ W}$. The noise data was linearly fitted with a slope of 1.48 kHz/W corresponding to a generated NSD of $(110\pm 4) \mbox{ kHz/nm}$ at our maximum internal efficiency of 72\%.}
\label{fig:conversionandnoise}
\end{figure}

\begin{figure}[!htbp]
\centering
\includegraphics[width=8.8cm]{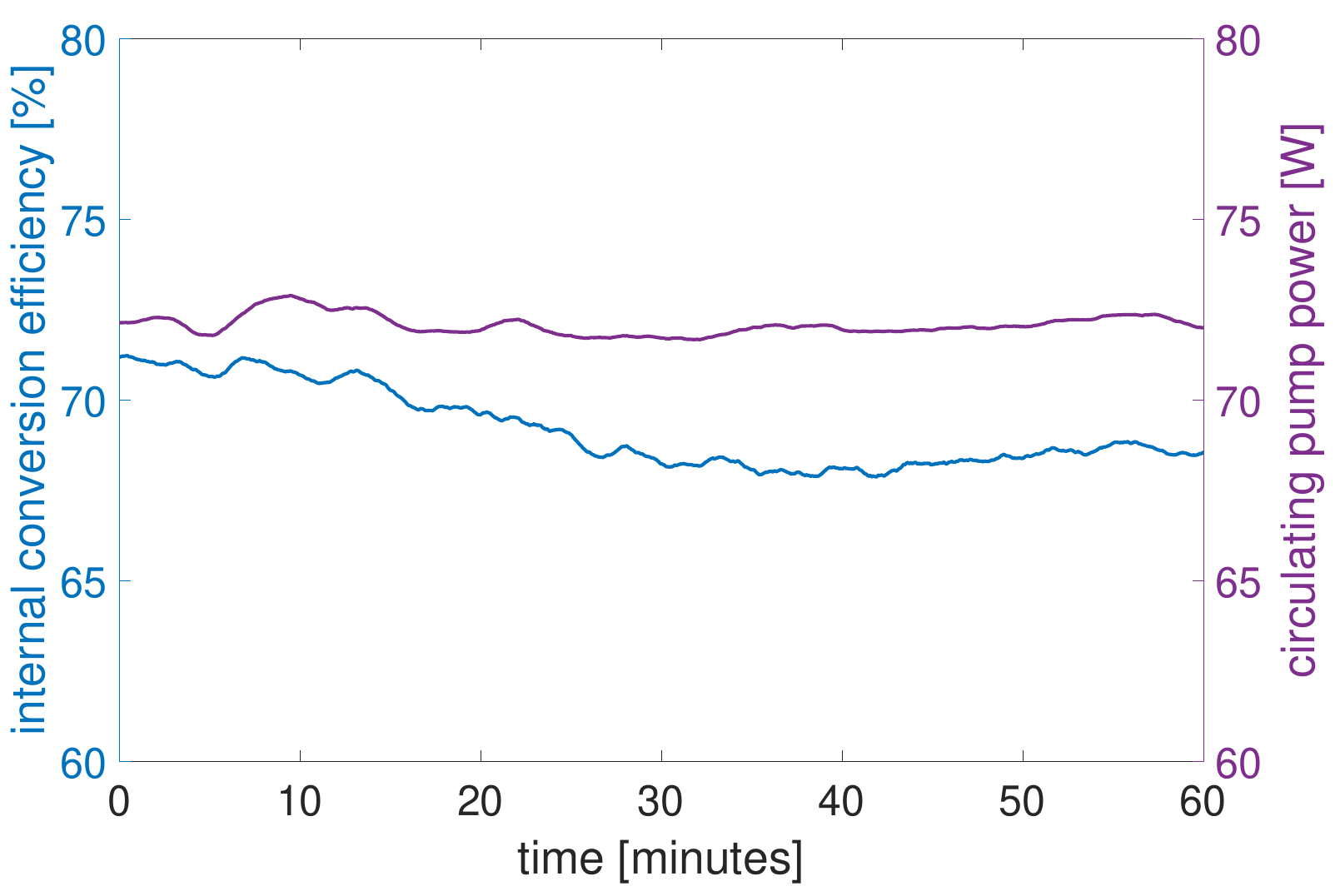}
\caption{The measured stability of the depleted weak probe (blue curve) and the circulating pump power (purple curve) over 60 minutes. The mean values of the conversion efficiency and the circulating pump power were (69.2$\pm$ 1.2)\% and (72.1 $\pm$ 0.3) W respectively. The small degradation of the conversion efficiency over time was due to a slow drift of the wavelength of the red probe laser.}
\label{fig:longterm}
\end{figure}

\begin{figure}[!htbp]
\centering
\includegraphics[width=8cm]{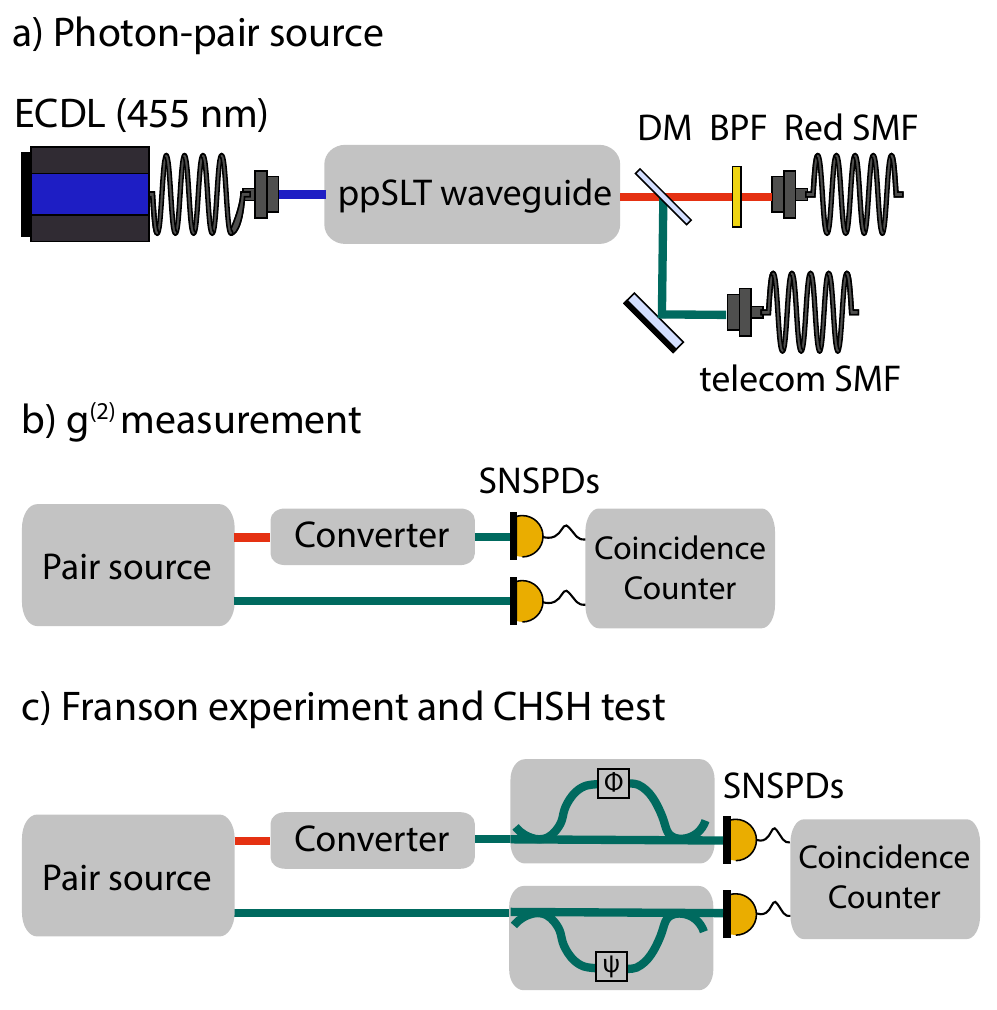}
\caption{a) Photon pair source based on a ppSLT waveguide and a blue ECDL b) Setup for the $g^{(2)}_{s,i}$ measurement including a conversion step c) Setup with asymmetric Mach-Zehnder interferometers before the detectors in order to perform a Franson experiment and the CHSH inequality test}
\label{fig:setup_additonal}
\end{figure}

Figure~\ref{fig:conversionandnoise} shows the measured internal conversion efficiency and the generated noise spectral density NSD as a function of the circulating pump power. The collection bandwidth of the converter was determined by the monochromator formed by the holographic-transmission grating and the aperture of the single mode fiber, with an experimentally measured FWHM bandwidth of $0.87$ nm at 1587 nm. At the maximum circulating pump power of $(74.5\pm 0.3)\mbox{ W}$ the internal (external) conversion efficiency was 72\% (33\%) while generating $(110\pm 4) \mbox{ kHz/nm}$ ($(45\pm 2)\mbox{ kHz/nm}$) noise. This is a 5-fold reduction in NSD over that of the best state-of-the-art single-step converter \cite{dreau}.
\\The measured noise counts were verified with both SNSPDs. We currently attribute the remaining noise in our converter to be largely due to parasitic SPDC. This is due to the noise spectrum exhibiting characteristic spectral features which can be tuned with crystal temperature, pump polarisation and pump wavelength. Detailed investigations on this will be presented soon in a future publication.
\\The internal conversion efficiency was determined via a power depletion measurement of a weak red probe, and the external conversion efficiency was determined by directly measuring the ratio between the mean photon number of an injected red weak probe and that of the converted telecom light coupled into a SMF.
\\The measured internal conversion efficiency in Figure~\ref{fig:conversionandnoise} suggests a $P_{max}=177 \mbox{ W}$ while Eqn.~\ref{eq:DFG} predicts a considerably lower value, assuming a typical value for $d_{eff}$. We attribute this discrepancy to imperfect mode matching and alignment, and possibly spatial deformation of the modes through conversion. Accurate estimation of the generated noise spectral density, $NSD_{gen}$ required a careful accounting of the efficiencies in the experiment. The data presented in Figure~\ref{fig:conversionandnoise} was calculated with
\begin{gather}
    NSD_{gen}= \frac{N_{meas}}{\delta \lambda \cdot \eta_{col}\cdot \eta_{det}},
    \label{eg:NSD}
\end{gather}
where $\eta_{col}= 33\%/72\%\approx 46 \%$ was our single-mode fiber collection efficiency of converted light/noise generated in the crystal. It resulted from the fiber coupling efficiency (72\%) and the transmission through optics (99\%) and filters (BPF: 92\%, TG: 70\%). Note that a reduction in these losses only by one half using optimized optical components would already result in a total system efficiency $>$50\%, while their further optimization together with a doubling in circulating pump power would make $>$80\% total conversion efficiency achievable.
\\Due to a thermal self-tuning effect, no active stabilization was required to keep the pump on resonance and the conversion stable over hours \cite{mitchell,mitchell2}. The stability of the circulating pump power and of the converted signal over 60 minutes can be seen in Figure~\ref{fig:longterm}. The observable small degradation of the conversion efficiency was due to a slow drift of the wavelength of the red probe laser used for the depletion measurement.\\
The converter can be used in a bidirectional manner due to the presence of the forth- and back-traveling pump wave in the cavity. This in principle allows the converter to be used simultaneously for two independent conversion tasks, e.g. for the conversion of polarisation qubits via spatial multiplexing \cite{ramelow,becher}.
\\To highlight the suitability of our converter for quantum applications, we demonstrate the preservation of non-classical correlations by the conversion process by interfacing a photon pair source based on SPDC (compare Figure \ref{fig:setup_additonal}). A source of single photons from NV centers in diamond was unfortunately not available at that time to test the converter. The 35 mm long ppSLT laser-written depressed-cladding waveguide source \cite{laserwritten} was pumped via a CW external cavity diode laser (ECDL) at 455 nm, generating pairs at 637 nm and 1587 nm.\\
Interfacing the signal photon with the converter, a peak value of the normalized cross-correlation function of $g^{(2)}_{s,i}(\tau=0)=310.45\pm 0.25$ was obtained. This measurement violates a Cauchy-Schwarz inequality by more than 1200 standard deviations, indicating that the non-classical character of the photon pair source has been preserved through conversion \cite{ppKTPWGnoise,walls}. 364 kHz of singles from the converter and 835 kHz singles on the other detector were measured. With a coincidence window of 1 ns the coincidence rate was 18.5 kHz. See Appendix \ref{app:g2} for additional information.
\\To further highlight the performance of our converter, we performed a Franson experiment \cite{franson} by inserting two asymmetric Mach-Zehnder interferometers (AMZIs) into the paths of the idler and the now down-converted signal photons. We observe raw fringe visibilities of up to $(98.2 \pm 0.1)\%$, far exceeding the classical limit of 50\%. See Table 1 in Appendix \ref{app:franson} for the visibilities of the different port combinations.
\\Additionally, a more rigorous CHSH-type chained Bell inequality test was carried out to demonstrate the preservation of time-energy entanglement by the converter. It was pointed out in \cite{chainedbell2} that standard CHSH inequalities cannot be violated in a Franson experiment due to the required post-selection increasing the classical bound beyond the quantum bound. This motivated the development of chained versions of the CHSH-inequality, allowing for a violation in such experiments, but demanding high fringe visibilities exceeding 94.63\%. Performing a Bell-test using the 10 term inequality in \cite{chainedbell2} we obtained an experimental Bell-parameter of 
\begin{gather}
S_{measured}=9.282\pm 0.017.
\end{gather}
Considering $S_{LHV} = 9$ this violates the 10 term CHSH-type chained Bell inequality by $16\sigma$. The quantum mechanical prediction is $S_{QM} = 9.511$. See Table 2 in Appendix \ref{app:bell} for the measurement results for the 6, 8, 10 and 12 term inequalities.

\section{Performance}
\begin{figure}[!htbp]
\centering
\includegraphics[width=9cm]{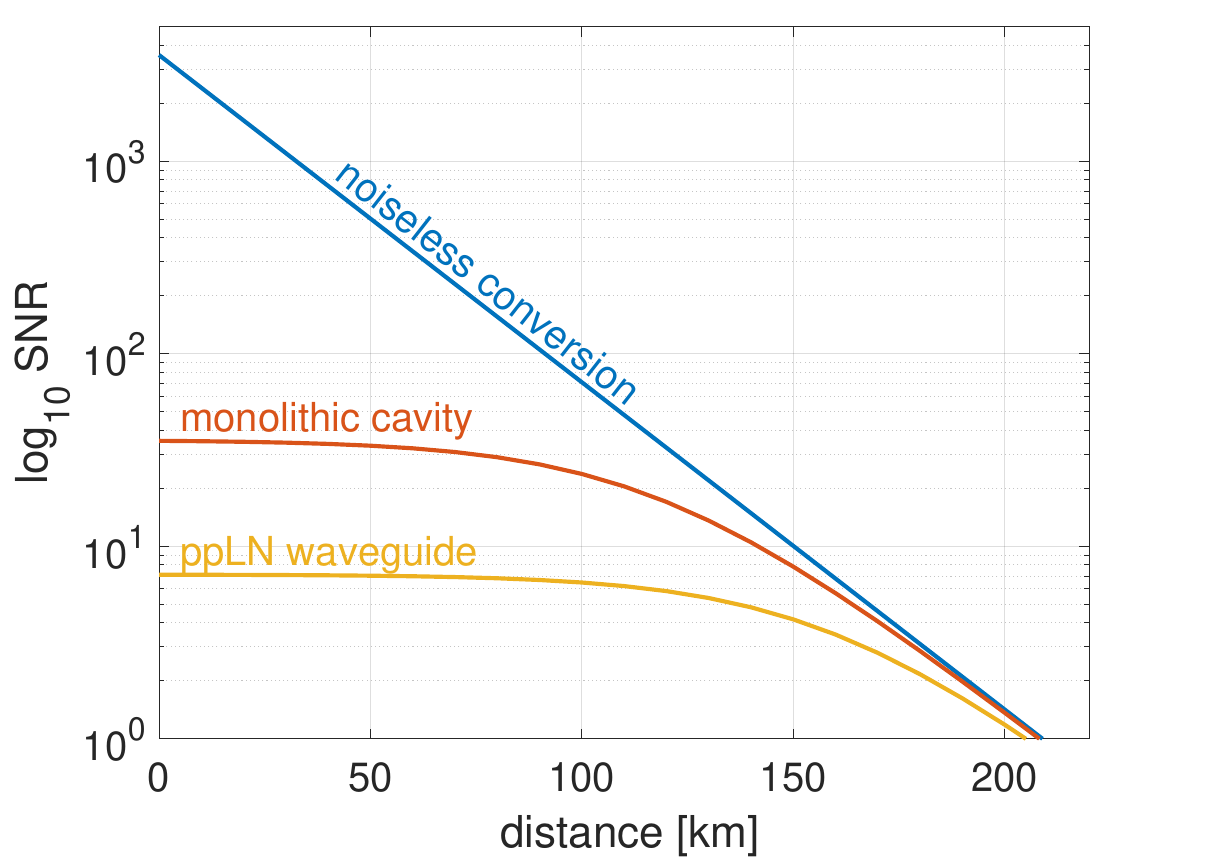}
\caption{Estimated signal-to-noise ratio (SNR) in logarithmic scale as a function of fiber transmission distance for noiseless conversion (blue), the conversion noise level of the presented monolithic cavity (red) and the conversion noise level of the so far best single-step ppLN waveguide converter \cite{dreau} (orange).}
\label{fig:SNR}
\end{figure}

To best illustrate the performance of our converter for a potential quantum network based on NV centers in diamond, we show in Figure~\ref{fig:SNR} the signal-to-noise ratio (SNR) as a function of fiber transmission distance $x$ compared to that of a noiseless converter as well as that of the current state-of-the-art ppLN waveguide converter \cite{dreau}. The simple model used for this is given by
\begin{gather}
    \mbox{SNR(x)}=\frac{\eta_{ext}\cdot R_{NV}\cdot\exp(-\alpha_t x)}{N_{C}\cdot\exp(-\alpha_t x)+N_{D}}.
    \label{eq:SNR}
\end{gather}
Here $\eta_{ext}$ includes all losses not scaling with distance: the external conversion efficiency of 33\%, narrowband filter ($\mbox{FWHM}=5 \mbox{ pm}$) losses of 50\% and the detection efficiency of 80\%. $R_{NV}$ is the rate of the single photons from NV centers in diamond, $N_C$ the noise rate of the converter and $N_D$ the dark noise rate of the detector. $\alpha_t$ is the loss per distance in the telecommunication fiber. This extends the analysis of "point-to-point links" for example in \cite{vanloock}, where converter noise and detector dark counts are not considered. In the here presented example, for all cases the same $ R_{NV} = 26.5 \mbox{ kHz}$, $\eta_{ext} \approx 0.15 $, $\alpha_t=-0.17\mbox{dB/km}$ and $N_D = 1 \mbox{ Hz}$ were assumed.
Without conversion of the 637 nm photon to telecom wavelength, the losses are so strong that the SNR would effectively drop below one within a few kilometers. In contrast, quantum frequency conversion not only allows for distances of up to about 150 km, but when compared to the state-of-the-art ppLN waveguide converter \cite{dreau}, the bulk ppKTP converter presented here offers a significant constant advantage that is maintained up to 100 km. Our assumed 1 Hz of detector noise starts limiting the overall SNR for the presented converter in the range of 100-150 km, diminishing the advantage of further suppressing converter noise. Above 150 km the SNR for all three scenarios drop significantly and quantum repeater schemes become necessary \cite{vanloock} to maintain suitably large rates. Note that, because the dark counts of the detector are assumed to be 1 Hz for noiseless conversion the SNR is equal to the expected rates per second which can be recorded at a certain distance.
\\While at large distances, the performance of our converter asymptotes towards that of an ideal converter, over shorter distances -- ranges  associated with metropolitan or intra-city links -- a considerable performance gap remains. Closing this gap requires foremost a further reduction in the noise floor of the converter, and secondarily, improvements to the efficiency (both internal and external).\\
\section{Conclusions}
While bulk ppKTP has facilitated a considerable improvement in the noise floor over the existing reported values, there still remains significant room for improvement for single-step converters. Detailed measurements of the noise at the target wavelength here lead us to attribute it near-exclusively to spurious SPDC, namely the unwanted down-conversion of the strong 1064 nm pump to a signal photon at the target wavelength and an idler in the mid-IR. One might think shifting from quasi-phase matching to birefringent phase matching would therefore offer a solution, but in doing so one both loses access to the comparatively strong d$_{33}$ tensor element and introduces double refraction, demanding much higher (an order of magnitude more) pump powers and experimental complexity and lower scalability, all the while limiting the intrinsically attainable maximum conversion efficiency significantly below 100\%. Instead, approaches that focus on direct suppression of the unwanted SPDC processes in periodically poled crystals could involve resonantly enhancing the signal field, which would dramatically lower the required circulating pump power. Alternatively or additionally, one could realize schemes with strong absorption of, or utilizing anti-resonant cavity suppression of, the parasitic mid-IR idler field.\\
In conclusion, we have presented an efficient low-noise quantum frequency converter based on a monolithic bulk ppKTP cavity suitable for the conversion of single photons from nitrogen-vacancy centers in diamond to telecommunication wavelength. Together with the compact and self-stabilizing design and requiring only cost-effective and off-the-shelf commercial pump lasers, our QFC-platform represents a viable technological ingredient for improved scalability of future quantum networks.
\newpage
During the review process of this publication, the authors became aware of a relevant preprint on low-noise conversion using an unpoled KTA crystal \cite{geus}.

\section*{Acknowledgements} We thank Martin Jutisz, Max Tillmann, Florian Kaiser and Janik Wolters for helpful discussions and technical support.
\section*{Funding Information} Funded by the BMBF Germany within the project QR.X.
\section*{Disclosures} The authors declare no conflicts of interest.

\appendix

\section{On the $g^{(2)}_{s,i}$ measurement}\label{app:g2}

\begin{figure}[!htbp]
\centering
\includegraphics[width=8.3cm]{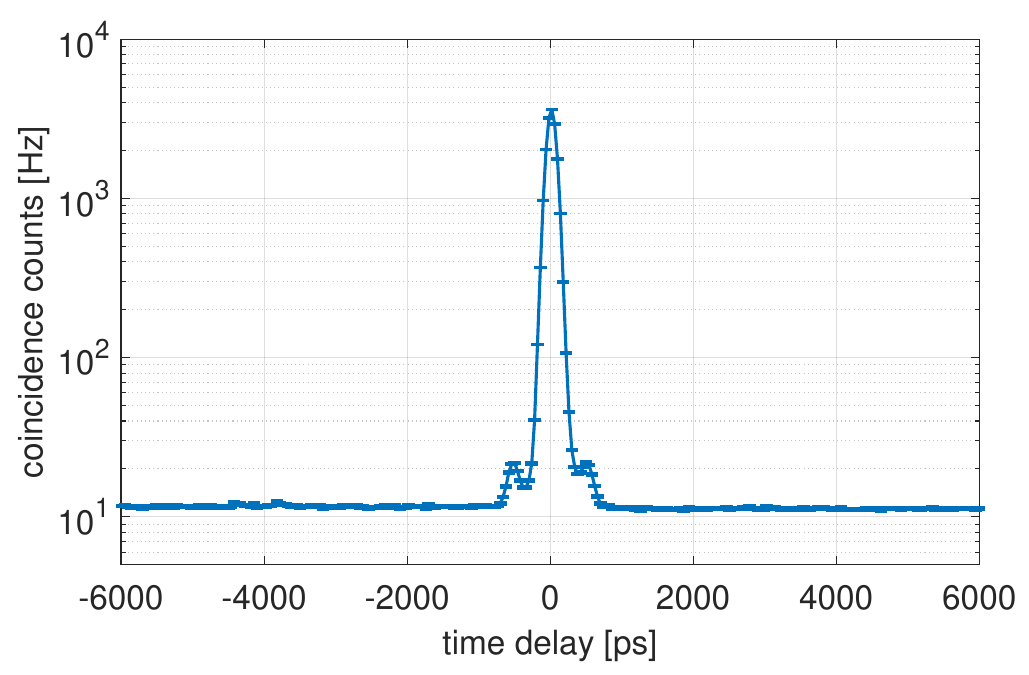}
\caption{Measured unnormalized $G^{(2)}_{s,i}$-cross-correlation function of the photon pairs after the conversion with error bars of the standard errors.}
\label{fig:g2}
\end{figure}

Figure \ref{fig:g2} shows the measured unnormalized cross-correlation function after the down-conversion of the signal photon. From this a peak value of the normalized cross-correlation function $g^{(2)}_{s,i}(\tau=0)=310.45\pm 0.25$ was obtained. The small sidelobes evident in the Figure at $\pm 500\mbox{ ps}$ from the centre peak are attributed to residual reflections in the detection electronics.

\newpage

\section{On the Franson experiment}\label{app:franson}
\begin{figure}[!htbp]
\centering
\includegraphics[width=8.5cm]{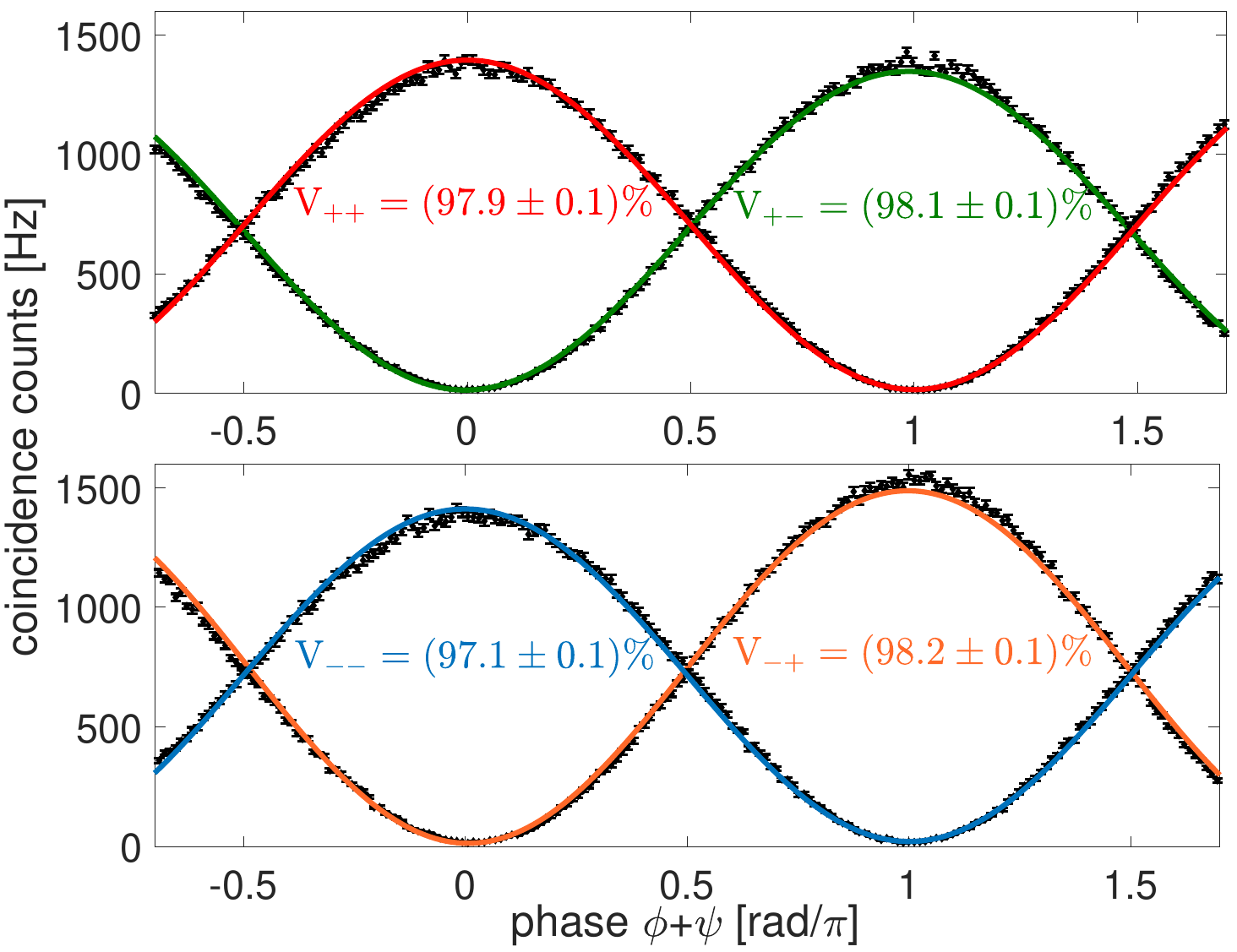}
\caption{Measured raw coincidences in the Franson experiment with postselection over detuned phase $\psi+\phi$ for all four port combinations with error bars of the
standard errors. $cos^2(x)$ functions where fitted to extract the visibility. A raw fringe visibility of the fits of about 98\% was reached (see Table \ref{table:vis}).}
\label{fig:franson}
\end{figure}

Figure \ref{fig:franson} shows the measured raw coincidences in the Franson experiment. In Table \ref{table:vis} all visibilities of the different output port combinations from the Franson experiment are listed. Raw (corrected) visibilities are without (with) the subtraction of the accidentals background.

\begin{ruledtabular}
\begin{table}[htb]
 \centering \caption{Visibilities in the Franson experiment}
\begin{tabular}{cccc}
port & raw visibility  & corrected visibility \\
\hline
    $++$ & $(97.9\pm 0.1)\%$ & $(98.5 \pm 0.1)\%$ \\
    $+-$ & $(98.1\pm 0.1)\%$ & $(98.7 \pm 0.1)\%$ \\
    $--$ & $(97.1 \pm 0.1)\%$ & $(97.7 \pm 0.1)\%$ \\
    $-+$ & $(98.2 \pm 0.1)\%$ & $(98.8 \pm 0.2)\%$ \\
\end{tabular}
\label{table:vis}
\end{table}
\end{ruledtabular}

\section{On the chained Bell inequalities}\label{app:bell}

A CHSH-type chained Bell inequality test was carried out to demonstrate the preservation of time-energy entanglement by the converter. It was pointed out that standard CHSH inequalities cannot be violated in a Franson experiment due to the required postselection increasing the classical bound beyond the quantum bound \cite{chainedbell1,chainedbell2}. This postselection loophole, would even allow for faking the violation of the CHSH inequality with classical light \cite{hacking} and thus cannot serve as an entanglement witness \cite{witness,terhal} without making additional assumptions.\\
Nonetheless, chained versions of the CHSH inequality were developed \cite{wringing,chainedbell3,chainedbell4}, allowing for a violation in experiments, but requiring very high fringe visibilities exceeding 94.63\%. A postselection-loophole-free Bell violation with genuine time-bin entanglement has been performed recently \cite{chainedbell3}. The chained CHSH inequality for N measurement settings and 2N terms is given by \cite{chainedbell4}
\begin{gather}
    S^N \leq S_{LHV}^N,\label{eq:inequality1}\\
    S^N = \braket{A_N B_N} + \sum_{k=2}^N\left[ \braket{A_k B_{k-1}}+\braket{A_{k-1} B_k}\right]-\braket{A_1 B_1}.
    \label{eq:inequality2}
\end{gather}
The upper bound for local hidden-variable theories is $S_{LHV}^N=2N-1$ while quantum theory predicts $S_{QM}^N=2N\cos\left(\frac{\pi}{2N}\right)$. The strongest violation by quantum mechanics is given for a relative angle of $\theta=\phi+\psi=\frac{\pi}{2N}$. The critical visibility which has to be surpassed experimentally is given by $V_{crit}=S^N_{LHV}/S^N_{QM}$. See Table \ref{table:bell} for all results.

\begin{ruledtabular}
\begin{table}[htb]
 \centering \caption{Chained Bell inequality violations}
\begin{tabular}{cccccc}
N & $S^N_{LHV}$ & $S^N_{QM}$ & $V_{crit}$ & $S^N_{measured}$ & Violation \\
\hline
    $2$ & $3$ & $2.828$ & $>100\%$ & $2.755\pm 0.020$ & $-$\\
    $3$ & $5$ & $5.196$ & $96.23\%$ & $5.039\pm 0.017$ & $2\sigma$\\
    $4$ & $7$ & $7.391$ & $94.71\%$ & $7.270\pm 0.017$ & $15\sigma$\\
    $5$ & $9$ & $9.511$ & $94.63\%$ & $9.282\pm 0.017$ & $16\sigma$\\
    $6$ & $11$ & $11.59$ & $94.90\%$ & $11.335\pm 0.018$ & $18\sigma$\\
\end{tabular}
\label{table:bell}
\end{table}
\end{ruledtabular}

In our experiment, we did not have a closed-loop phase control with stabilisation light \cite{chainedbell4}. Keeping a phase setting stable was working well but the repeatability of setting a phase was not ideal. This is due to the properties of an open-loop piezo positioning-system. Since each phase measurement operator $A_j$ and $B_j$ appears twice in the chained Bell inequality each phase setting can just be kept to record two expectation values. This may lead to unideal relative angles $\theta$ and thus decrease the violation but it is allowed by Formulas \ref{eq:inequality1} and \ref{eq:inequality2}. Not choosing the phase settings pairwise (but independently) as dictated by these equations on the other hand would increase $S_{LHV}$. Only the last term has one phase setting of the first term and thus demands repeatability. In our measurements, we took care of the lack of repeatability by recording each chained Bell inequality 10 times and averaging over these trials. Further the expectation values in the experiment are given by $E=\frac{R_{++}- R_{+-}- R_{-+}+R_{--}}{R_{++}+ R_{+-}+ R_{-+}+R_{--}}$ where R are the rates again and the first index AMZI 1 and the second index AMZI 2 with ports "$+$" and "$-$" each. We just had two instead of four detectors available to monitor the four output ports of the interferometers. Thus we did assume $R_{++}=R_{--}$, $R_{+-}=R_{-+}$ and $R_{+-}=R^{\pi}_{++}$ \cite{chainedbell4}. Also, no fast random switching of the phases at the timescale of 1.25 GHz was employed here. The fastest frequency which would in principle be allowed by the piezos is 50 kHz.

%\bibliography{bib}

\end{document}